\documentclass[preprint,eqsecnum,aps,showpacs,tightenlines,
nofootinbib,showkeys,byrevtex,amsfonts]{revtex4}

\usepackage{graphicx}%
\usepackage{dcolumn}
\usepackage{amsmath}
\usepackage{epsfig}
\usepackage{latexsym,pslatex,bm}

\def\bfq{{\bm q}} 
\def\bfk{{\bm k}} 
\def\bfp{{\bm p}}

\newcommand{\gsim}{\raisebox{-0.7ex}{$\stackrel{\textstyle >}{\sim}$ }}
\newcommand{\lsim}{\raisebox{-0.7ex}{$\stackrel{\textstyle <}{\sim}$ }}
\def\darr#1{\raise1.5ex\hbox{$\leftrightarrow$}\mkern-16.5mu #1}
\def\){\right)} 
\def\({\left(} 
\def\]{\right]} 
\def\[{\left[}

\def\Align{&=&}

\def\npdg{$np\rightarrow d\gamma$}

\def\si{{}^1\kern-.14em S_0}
\def\siii{{}^3\kern-.14em S_1}
\def\diii{{}^3\kern-.14em D_1}

\def\pone{{}^3\kern-.14em P_1}
\def\pzero{{}^3\kern-.14em P_0}
\def\ptwo{{}^3\kern-.14em P_2}

\def\nrcpt{NR\raise.4ex\hbox{$\chi$}PT\ }

\def\ltap{\ \raise.3ex\hbox{$<$\kern-.75em\lower1ex\hbox{$\sim$}}\ }
\def\gtap{\ \raise.3ex\hbox{$>$\kern-.75em\lower1ex\hbox{$\sim$}}\ }

\newcommand{\eqn}[1]{\label{eq:#1}}
\newcommand{\refeq}[1]{(\ref{eq:#1})}

\newcommand{\beq}{\begin{eqnarray}}
\newcommand{\eeq}{\end{eqnarray}}

\newcommand{\mcal}[1]{{\mathcal #1}}

\newcommand{\makefigR}[4]{\begin{figure}[!t] 
	 		 \includegraphics[height=#3]{#2}%
                           \caption{#4 } 
			   \label{#1}
                         \end{figure}}

\def\Journal#1#2#3#4{{#1} {\bf #2}, #3 (#4)}

\def\NPB{{\em Nucl. Phys.} B}
\def\NPA{{\em Nucl. Phys.} A}

\def\PLB{{\em Phys. Lett.} B}
\def\PRL{\em Phys. Rev. Lett.}

\def\PRC{{\em Phys. Rev.} C}

\begin{document}
\bibliographystyle{unsrt}
\preprint{TRI-PP-01-13}

\title{Quartet $S$-wave $p$-$d$ scattering in EFT}

\author{Gautam Rupak
}
\email[Corresponding author. Email: ]{grupak@triumf.ca}

\author{Xin-wei Kong}

\affiliation{
TRIUMF, Vancouver, B.C.\\
Canada V6T 2A3
}


\begin{abstract}
We present a power counting to include Coulomb effects in the three-nucleon
system in a low-energy pionless effective field theory (EFT).
With this power counting,
the quartet $S$-wave
proton-deuteron elastic scattering amplitude is calculated.
The calculation includes
next-to-leading order (NLO) Coulomb effects and next-to-next-to-leading order
(N$^2$LO) strong interaction effects, with an estimated theoretical error of
$\sim 7\%$. The EFT results agree with potential model calculations and phase
shift analysis of experimental data within the estimated errors.
\end{abstract}

\pacs{11.80.Et; 11.80Jy; 13.40Ks; 13.75.Cs;
21.30.-x; 21.30.Fe; 21.45.+v; 25.10.+s; 25.40.Cm; 25.45.De}
\keywords{proton-deuteron scattering, Coulomb effects,
effective field theory, three-body system.}
\maketitle

\begin{section}{Introduction}

In the last few years there have been a spur of activities involving nuclear
effective field theories (EFT)~\cite{spur2,BBHPS2000}.
Use of EFT in nuclear physics is not a new idea.
Chiral perturbation theory ($\chi$PT) is
an EFT that has been phenomenologically successful in the one nucleon
sector. The next break-through came with Weinberg's work on extending
$\chi$PT techniques to the two-nucleon systems~\cite{Weinberg1,bira}.
Following Weinberg's work, EFTs differing in expansion
parameter, dynamical content, regularization procedure, etc.,
were discovered, rediscovered and developed further.
The initial period was devoted mostly to understanding these EFTs by
looking at nucleon-nucleon elastic scattering
amplitudes~\cite{mehen,RS,CohHan99,FMSep}.
From these studies,
two alternate formulations of the nuclear EFT emerged as relevant for
calculating nuclear cross sections at low-energies. The first EFT is based on
Weinberg's original proposal where one expands the two-nucleon potential in
perturbation, and then solves the Schr\"odinger equation with this potential
for the nucleon wave function. This EFT includes nucleons, pions and
photons as dynamical degrees of freedom~\cite{bira,Epel}.
In the second EFT, one expands the
scattering amplitude directly in perturbation, through the use of Feynman
diagrams. Here one includes only nucleons and photons as dynamical degrees of
freedom~\cite{bira2,threebod,gegelia,CRSa}. This theory is applicable at
momenta $p$ smaller than the pion mass
$m_\pi$, which is quite suitable for various nuclear reactions relevant for
nuclear astrophysics. In this paper, we will work with the pionless EFT as
described in Ref.~\cite{CRSa}.

Now, we briefly describe the nuclear EFT procedure here.
The interested reader should look at the comprehensive and up to date review in
Ref~\cite{BBHPS2000} for details. EFT is a useful tool in the study of
physical processes
with a plethora of clearly separated physical scales.
This is generally the case in the few-nucleon sector at low-energies, where for
example, the deuteron binding momentum $\gamma\sim 45$ MeV is smaller than the
pion mass $m_\pi\sim 140$ MeV, which in turn is smaller
than the nucleon mass $M_N\sim 1000$ MeV, etc. EFT
provides a natural scheme for separating the short distance physics from the
long
distance effects. At external momenta $p\sim m_\pi$, pion mass $m_\pi$
sets the long distance scale and nucleon mass $M_N$, heavier meson masses set
the short distance scale. Currently, we are interested at momenta
$p\sim \gamma$ smaller than the pion mass. Thus we construct an appropriate
low-energy non-relativistic
EFT where pion effects are part of the high-energy physics.
The
strong interaction of the nucleons is then described by the most general set of
multi-nucleon-photon local operators $\mcal O_i$,
respecting the low-energy symmetries,
in an EFT Lagrangian:
\beq\eqn{EFT_L}
\mcal L(x)\Align \sum_i C_i\mcal O_i(x)\ ,
\eeq
where the effects of the pions and other heavier dynamical particles that
 were ``integrated out'' of the theory are encoded in the ``high-energy''
coefficients $C_i$'s. The dimensionful couplings $C_i$'s are assumed to depend
only on the high-energy scales $m_\pi$, $m_\rho$, etc., $\sim\Lambda$
and they are determined
from a fit to experimental data. To make any meaningful
prediction with the Lagrangian in Eq.~\refeq{EFT_L}, one develops
unambiguous power counting rules that determine which operators in
Eq.~\refeq{EFT_L} are most important and which are not. Typically only a few
operators are required. Thus one can predict a multitude
of processes once a few
unknown couplings are determined from a few low-energy experiments.
In addition to this,
one also requires power counting rules to estimate loop diagrams that describe
quantum effects. With these power counting rules one finally expresses all
physical observables
in a perturbative expansion of local operators and loops,
where the expansion parameter is expected to be $Q/\Lambda$ with $Q\sim p$,
$\gamma$ and $\Lambda\sim m_\pi$.
The perturbative description of
the low-energy physics then allows a systematic estimation of errors at any order in the perturbation.

A word about the high-energy cut-off $\Lambda$ is appropriate here.
A priori one cannot
determine the
exact value of $\Lambda$ in an EFT calculation. This would require
complete knowledge of the high-energy couplings $C_i$'s and
all the loop effects, which one does not have. $\Lambda$ can be empirically
estimated from EFT calculations by comparing the contributions from different
orders in the perturbation.
$\Lambda$ is process dependent, however, it is found to
be $\sim m_\pi$ from various other EFT calculations.

Recently, effort has been directed towards applications of
pionless nuclear EFT, especially in the two-nucleon
systems involving external
currents~\cite{CRSa,CRSb,BC2001,BCK2000,KFpp,KFnu,rupak99,GR2000}.
Progress has been made and these calculations have added much to our
understanding. In terms of accuracy, some of these calculations are at least as
good as traditional model calculations.
On the other hand, the three-body
EFT calculations have been so far confined only to the
$n$-$d$ system~\cite{threebod,BG99,GBG99,FG2001,HM2001,Bedaque:2002yg}.
In the $n$-$d$ doublet
channel there is a three-nucleon contact operator at
leading order (LO), whereas in
the quartet channel such three-nucleon operators do not contribute
up to next-to-next-to-leading order (N$^2$LO).
 These
calculations reproduce available experimental data where applicable within the
theoretical errors assumed.

Application of EFT to study $p$-$d$ scattering is a natural extension of the
work carried out so far.
An interesting aspect of calculating
$p$-$d$ scattering amplitude
would be to study its analyzing power $A_y$. This might shed some light
on the long standing $A_y$ puzzle.
$p$-$d$ scattering will also play an important role in
EFT calculations of low-energy processes
such as
$pd\rightarrow\gamma\ ^3$He, $dd\rightarrow n\ ^3$He, etc.,
that are important input for primordial light element prediction in big-bang
nucleosynthesis~\cite{BNTT}.
Triton beta decay and certain neutrino-deuteron scattering processes
receive contributions from the same axial current
operators with undetermined coupling $L_{1,A}$~\cite{BC2001,OnBetaDecay}.
The unknown $L_{1,A}$ is one of the large sources of uncertainty in the EFT
neutrino-deuteron scattering calculations.
One could in principle determine $L_{1,A}$ from
triton beta decay data with reasonable accuracy.
Understanding Coulomb effects in the $p$-$d$ system will be crucial for the
EFT triton beta decay calculation.

The pionless EFT would be an ideal tool to calculate these low-energy
cross sections in a model-independent
way and to, possibly, reduce the theoretical errors as has been done for
the \npdg~process~\cite{rupak99}. However, all these require a systematic
handling of the Coulomb photons in the many-nucleon system as has been
done for the purely strong
interaction. Elastic $p$-$d$ scattering in the quartet channel provides
a unique situation to study Coulomb effects in many-nucleon systems. It is
complicated enough in the sense that it involves strong interactions and
Coulomb
effects in a few-nucleon system. On the other hand, one does not have to worry
about three-nucleon forces in the quartet channel as shown in
Ref.~\cite{threebod,BG99,GBG99}.

One of the primary goals of this calculation is to establish and understand the
EFT power counting for the dominant Coulomb corrections at low-momentum.
This is crucial for the three-nucleon EFT calculations involving more than the
$n$-$d$ system. EFT might play an important role in precision calculations of
inelastic three-nucleon processes involving external currents.
We account
for the infrared divergent Coulomb contributions and
 as a first step reproduce in EFT formulation, potential
model results that have been known for decades.
The power counting for the strong interaction is well
established~\cite{KSW1,bira2,threebod,CRSa}
and it has been successfully applied to calculate
various two and three-nucleon
processes~\cite{spur2,CRSa,CRSb,BC2001,BCK2000,KFpp,KFnu,rupak99,GR2000,
threebod,BG99,GBG99,FG2001,HM2001}.
First, we recapitulate the strong interaction power counting, ignoring the
Coulomb corrections in Section~\ref{JustStrong}.
Our calculations will closely follow the power
counting in Ref.~\cite{KSW1,CRSa,bira2,gegelia}.
Then we develop the power counting for
the $p$-$d$ system interacting only through Coulomb photons in
Subsection~\ref{JustCoulomb}. In Subsection~\ref{StrongAndCoulomb}, power
counting for the $p$-$d$ system interacting through both the strong and Coulomb
interaction is developed. The phase shifts for the quartet $S$-wave
$p$-$d$ scattering
are considered in Section~\ref{phases}. We discuss the theoretical and
numerical errors in the calculation. A comparison with a potential model
calculation and phase shift analysis from experimental data is also
made. Finally, we present our conclusions in Section~\ref{conclusion}.
\end{section}
\begin{section}{Strong interaction power counting}
\label{JustStrong}
The strong
and Coulomb interactions in the $p$-$d$ system are described by the low-energy
Lagrangian~\cite{kaplan96,threebod,BG99,GBG99}:
\beq \eqn{dibaryon}
\mcal{L_{Nd}}\Align N^\dagger\[iD_0+\frac{{\bm D}^2}{2M_N}
-\frac{D_0^2}{2M_N}\] N
-d_i^\dagger\[w\(iD_0+\frac{{\bm D}^2}{4M_N}\)+\sigma_d\]d_i\nonumber\\
&{}&+\ y\[d_i^\dagger\(N^TP_iN\)+h.c\]+\cdots\ ,
\eeq
where ``$\cdots$'' represents higher dimensional operators with more
derivatives. The covariant derivative is:
\beq
D^\mu\Align\partial^\mu+ie\frac{1+\tau_3}{2}A^\mu\ ,
\eeq
and the $P_i$ matrices are used to project on to the $^3S_1$ state,
\beq
P_i&\equiv&\frac{1}{\sqrt{8}}\sigma_2\sigma_i\otimes\tau_3;\hspace{1in}
\mathrm{Tr}[P_i^\dagger P_j]=\frac{\delta_{i\ j}}{2}\ .
\eeq
The matrix $\sigma_i$ acts on the nucleon spin space and $\tau_i$ acts
on the nucleon isospin space.
$N$ is an isodoublet field representing the nucleons,
and $M_N\approx 938.92$ MeV is the isospin
averaged nucleon mass.
The auxiliary dibaryon field $d_i$ has
the same quantum numbers as a deuteron
and in the quartet channel $w=-1$. A Gau{\ss}ian  integration over the field
$d_i$ in the path integral, followed by a field redefinition, reduces
Eq.~\refeq{dibaryon} to the more familiar nuclear EFT Lagrangian with
four-nucleon interactions, etc.,~\cite{kaplan96,BG99,GBG99}.
The renormalized couplings $y$
and $\sigma_d$ can be determined from nucleon-nucleon scattering in the triplet
channel $^3S_1$~\cite{kaplan96,threebod,BG99,GBG99}.

In the EFT power counting, the expansion parameter is
$Q/\Lambda$~\cite{KSW1,CRSa}.
 All physical
observables are expressed as a perturbation in $Q/\Lambda$. The external
momenta $p$, the deuteron binding momentum $\gamma$ and the renormalization
scale $\mu$ are formally considered $\mcal O(Q)$ and $\Lambda\sim m_\pi$ for
this low-energy EFT. It is assumed in the power counting $y^2\sim 1/\Lambda$
 and $\sigma_d\sim Q$.

Compared to the LO, we will keep strong interaction
corrections up to $\mcal O(Q^2/\Lambda^2)$, i.e. N$^2$LO. Formally,
$m_\pi/M_N$ is taken to be $\mcal O(Q/\Lambda)$,
which is numerically consistent.
Thus, relativistic
corrections  which typically enter as $p^2/M_N^2$, $\gamma^2/M_N^2\sim
Q^2/M_N^2=Q^2/m_\pi^2\times m_\pi^2/M_N^2$ contribute at N$^4$LO, and we ignore
them here~\cite{CRSa,rupak99,GR2000}.

\begin{figure}[thb]
 \begin{center}
  \epsfig{figure=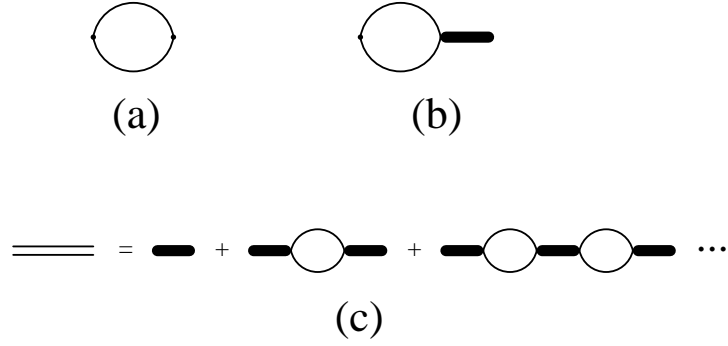, height=45mm}
 \end{center}
\caption{\it \protect The single lines represent nucleons,
 the filled double lines are dibaryons and double line dressed dibaryon.
The vertex couplings in diagram $\mathrm{(a)}$ are $1$.
All other
 vertex couplings are $y$.}
\label{PC1.fig}
\end{figure}
In this calculation we use dimensional regularization.
Some immediate consequences of the power counting are, after we integrate
over the energy component $q_0$ of a loop momentum, contracting a
 nucleon propagator:
\begin{enumerate}

\item The loop integration measure $\int d^3\bfq$ scales as $\mcal O(Q^3)$.

\item The nucleon propagator scales as $\mcal O(M_N/Q^2)$.

\item The dibaryon propagator is $-i/\sigma_d$ at LO.
Kinetic energy is
$\sim Q^2/M_N$ and it contributes at next-to-leading order (NLO) and higher.

\end{enumerate}

Equivalently, every nucleon propagator scales as $\mcal O(M_N/Q^2)$ and the
integration measure $\int dq_0 d^3\bfq$ scales as $\mcal O\[Q^5/(4\pi
M_N)\]$. We include a factor of $1/(4\pi)$ with every loop.
A closed nucleon momentum loop, Fig.~\ref{PC1.fig} (a), scales as
$\mcal O\[Q M_N/(4\pi)\]$.
Thus a two-nucleon
loop together with a dibaryon propagator,
Fig.~\ref{PC1.fig} (b), scales as: $\mcal O\[Q M_N/(4\pi)\]$ from the loop and
a factor of
$y^2\sim 1/\Lambda$
from the vertices and a factor of $i/\sigma_d\sim 1/Q$ from the dibaryon
propagator, which gives an overall factor of
$Q M_N y^2/(4\pi \sigma_d)\sim\mcal O(1)$.
Therefore, every diagram can be dressed up by an arbitrary number of
two-nucleon bubbles with a dibaryon propagator.

Now, the fully dressed deuteron propagator is given by the sum of diagrams in
Fig.~\ref{PC1.fig} (c). We get, for initial and final spin index
$\{i, j\}$ respectively,
\beq\eqn{deuteronLO}
i\Delta_{i\ j}(p_0,\bfp)\Align -i \frac{4\pi}{M_N y^2}\frac{\delta_{i\ j}}
{\frac{4\pi\sigma_d}{M_N y^2}-\mu +\sqrt{\bfp^2/4-M_N p_0-i\epsilon}}\cdot
\eeq
This dressed dibaryon propagator represents the deuteron propagator. It
is possible to include the dibaryon kinetic operator to all
orders~\cite{threebod,BS1,FG2001}.
This
allows one to trivially include N$^2$LO corrections due to the effective range
$\rho$ and greatly simplify the calculation.
However, the calculated $p$-$d$ scattering
amplitude will also include certain N$^3$LO and higher order effective range
corrections which should not be included in the strict perturbative sense. Note
that one could modify the power counting to formally count $\rho\Lambda$
as $\mcal O(\Lambda/Q)$ and
includ $\rho$ to all orders in perturbation, as done below~\cite{BS1}.
 We further add that only in the quartet channel (without the
three-body force) resumming the effective range to all orders reproduces 
the experimental results accurately~\cite{FG2001,HM2001,Bedaque:2002yg}. 
We find~\cite{GBG99}:
\beq\eqn{deuteronNLO}
i\Delta_{i\ j}(p_0,\bfp)\Align -i \frac{4\pi}{M_N y^2}\frac{\delta_{i\ j}}
{\frac{4\pi\sigma_d}{M_N y^2}-\mu +\sqrt{\bfp^2/4-M_N p_0-i\epsilon}
-\frac{4\pi}{M_N y^2}\(p_0-\bfp^2/(4 M_N)\)}\nonumber\\
&\equiv& i\Delta(p_0,\bfp)\delta_{i\ j}\ \cdot
\eeq
The two-nucleon scattering amplitude in the $^3S_1$ channel can now
be expressed in terms of the deuteron propagator as:
\beq
\eqn{3S1amp}
i\mcal A(k)\Align -y^2 i\Delta(p_0=k^2/M_N,\bfp=0)=\frac{4\pi}{M_N}
\frac{i}{k\cot\delta -ik}\ ,
\eeq
where the cotangent of the $S$-wave phase shift $\delta$ can be expressed
through the familiar effective range expansion:
\beq
k\cot\delta=-\gamma+\frac{\rho}{2}(k^2+\gamma^2)+w_2(k^2+\gamma^2)^2+\cdots\ ,
\eeq
where the deuteron binding momentum $\gamma=\sqrt{M_N B}$ with
the binding energy $B=2.224575(9)$ MeV~\cite{LA}.
$\rho=1.765(4)$ fm~\cite{nijmegen} is the
effective range, etc. From Eqs.~\refeq{deuteronNLO}
and~\refeq{3S1amp}, we get, ignoring shape parameter $w_2$, etc.,
\beq\eqn{deuteron}
y^2\Align-\frac{8\pi}{\rho M_N^2}\ ,\nonumber\\
\frac{4\pi\sigma_d}{M_N y^2}\Align \(\mu-\gamma\)
+\frac{\rho}{2}\gamma^2\ ,\nonumber\\
i\Delta(p_0,\bfp)\Align -i \frac{4\pi}{M_N y^2}\frac{1}
{-\gamma +\sqrt{\bfp^2/4-M_N p_0-i\epsilon}
-\frac{\rho}{2}\(\bfp^2/4-M_N p_0-\gamma^2-i\epsilon\) }\ ,
\eeq
and we also define the deuteron wave function renormalization factor:
\beq\eqn{wavefunction}
Z_d\Align\frac{1}{\partial_{p_0}(1/\Delta(p_0,\bfp))}
\Big|_{p_0=-\gamma^2/M_N;\  \boldsymbol{p}=0}
=\frac{8\pi}{M_N^2y^2}\frac{1}{\(\frac{1}{\gamma}-\rho\)}\ ,
\eeq
through the LSZ reduction procedure. The amputated amplitudes are multiplied by
factors of $\sqrt{Z_d}$ for every external deuteron propagator.

\begin{figure}[t]
 \begin{center}
  \epsfig{figure=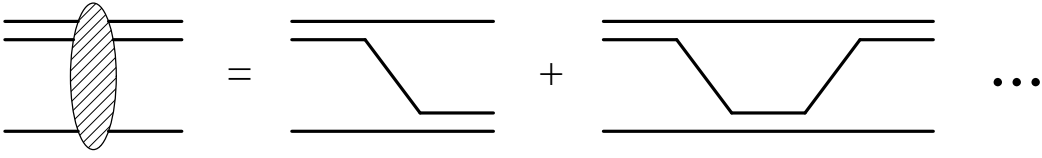, height=20mm}
 \end{center}
\caption{\it \protect Quartet $S$-wave
$p$-$d$ elastic scattering: the pinball diagrams.
The solid lines represent nucleons and double lines deuterons.
The deuteron-two-nucleon vertex coupling is $y$.}
\label{PD1.fig}
\end{figure}
Neglecting Coulomb effects, quartet $S$-wave $p$-$d$ scattering only
involves neutron exchange diagrams shown in Fig.~\ref{PD1.fig}. The tree level
amputated diagram involves two factors of $y$ from the vertices and a nucleon
propagator. Thus it scales as $y^2 M_N/Q^2$. A $n$-loop diagram in
Fig.~\ref{PD1.fig} would include an \emph{extra} factor of
$Q^{5n}/(4\pi M_N)^n$ from the
integration measure, a factor of $M_N^{2n}/Q^{4 n}$
from the nucleon propagators,
a factor of $\[4\pi/(y^2 M_N Q)\]^n$
from the dressed dibaryon propagators and a factor of
$y^{2n}$ from the vertices, for an \emph{overall extra} factor of
$\mcal O(1)$. Thus all the diagrams shown in
Fig.~\ref{PD1.fig} contribute to the scattering amplitude at LO.
In this theory higher order strong interactions include perturbative
corrections to the ratio $y^2/\sigma_d$ in Eq.~\refeq{deuteron}. As
mentioned earlier, we take effective range corrections into account to all
order in perturbation by using the relations in Eq.~\refeq{deuteron}. From
Fig.~\ref{PD1.fig}, we get for the purely strong half off-shell amputated
scattering amplitude:
\beq\eqn{S_Amp}
i\mcal T_{s}(\bfk,\bfp)\Align
\frac{-i M_N y^2}{{\bfk}\cdot{\bfp}+\bfk^2+\bfp^2 -M_N E_T}
\nonumber\\
&{}&-i 4\pi
\int {\frac{{\rm d}^3\bfq}{(2\pi)^3}\,}\frac{\mcal T_{s}(\bfk,\bfq)}
{-\gamma +\sqrt{3/4\bfq^2 -M_N E_T}-\frac{\rho}{2}\(3/4\bfq^2
-M_N E_T-\gamma^2\)}\nonumber\\
&{}&\hspace{.8in}
\times\frac{1}{\({\bfq}+{\bfp}/2\)^2+3/4\bfp^2-M_N E_T}\ ,
\eeq
where the incoming $\{ {\rm outgoing} \}$
deuteron carries momentum $\bfp\{\bfk\}$,
energy $\bfp^2/(4 M_N)-\gamma^2/M_N\{\bfk^2/(4 M_N)-\gamma^2/M_N+\epsilon\}$.
Similarly, the nucleon
carries momentum $-\bfp\{-\bfk\}$ and energy $\bfp^2/(2 M_N)
\{\bfk^2/(2M_N)-\epsilon\}$. Hence, the incoming deuteron and nucleon are
on-shell and the outgoing deuteron and nucleon are off-shell by an amount
$\epsilon$ and $-\epsilon$ respectively. The total center-of-mass energy $E_T$
of the $p$-$d$ system is $3\bfp^2/(4M_N) -\gamma^2/M_N$.
As in Ref.~\cite{BG99}, we set $\epsilon=(\bfk^2-\bfp^2)/M_N$ and then
$|\bfp|=|\bfk|$ puts all the external propagators on-shell to give the
on-shell amplitude.

\end{section}
\begin{section}{\protect Coulomb effects: \protect
\lowercase{$p$} and $Q$ counting}
\label{Coulomb}

Previously, when we neglected Coulomb interactions, it was assumed that the
external momentum $p$ and  the deuteron binding momentum $\gamma$ are of
similar size, i.e. $p\sim\gamma$. However, as known from non-relativistic
quantum mechanics,
Coulomb effects enter as $\alpha M_N/p$ and provide the dominant
contribution at low-energies.
Thus in estimating loop effects it is necessary to distinguish
between the two relevant physical scales $p$ and $\gamma$. In addition to the
expansion parameter $Q/\Lambda$
we introduce a new expansion parameter
$\sim p/(\alpha M_N)$. In the present non-relativistic theory there is no
pair-creation of either dibaryon or nucleon fields. However, a dibaryon field
does couple strongly to two nucleons.
Thus for low-energy $p$-$d$ scattering, all
the diagrams include at most one dibaryon field at a given time,
which can be put on-shell.
There are either one or three nucleon fields, one of which can be put
on-shell, the remaining two propagators being off-shell by an amount
proportional to the deuteron binding momentum $\gamma$. Thus every loop
integration have two dimensionful scales $p$ and $\gamma$, depending on
whether it involves a dressed dibaryon field or not. Thus, after integrating
over the time component $q_0$ and putting one nucleon on-shell, a loop integral
scales as some power of $p$ or $Q$ depending on weather we pick up the Coulomb
correction or the strong interaction effects.
We explain this in more detail in Appendix~\ref{powercounting}.

The previous power counting rules are modified as follows in the presence of
Coulomb effects:
\begin{enumerate}
\item{The loop integration measure $\int d^3\bfq$
scales as either $Q^3$ or $p^3$.}
\item{Every nucleon propagator scales as $M_N/Q^2$.}
\item{The dressed dibaryon propagator scales $Q/q^2$.
So depending on whether $d^3 q\sim Q^3$ or $d^3 q\sim p^3$,
the dibaryon propagator scales as $1/Q$ or $Q/p^2$.}
\item{Photon propagator scales as $1/Q^2$ or $1/p^2$,  depending on whether
$d^3 q\sim Q^3$ or $d^3 q\sim p^3$.}
\end{enumerate}
Rule $3.$ above is actually not different from the usual power counting where
one assumes $p\sim\gamma\sim Q$. When Coulomb photons are involved at low
momenta one needs to
generalize to the case $p\ll\gamma\sim Q$.
An immediate consequence of these rules is that in a loop with only nucleons,
all the momenta scale only as $Q$. These rules become clear when we
consider some typical Coulomb diagrams. Look at the examples in
Appendix~\ref{powercounting} as well.

\begin{subsection}{ Coulomb Ladder in EFT}
\label{JustCoulomb}

\begin{figure}[t]
 \begin{center}
  \epsfig{figure=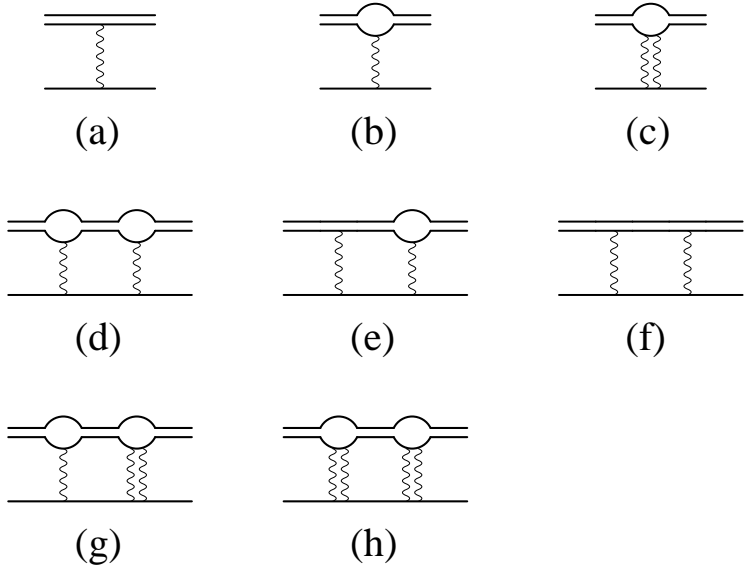, height=75mm}
 \end{center}
\caption{\it \protect Double lines: deuterons;
single straight line: nucleon fields; wavy lines: photons.
The deuteron-two-nucleon vertex coupling is $y$ and the photon-nucleon,
-deuteron coupling is $e$. }
\label{pd2.fig}
\end{figure}
It is easiest to start with diagrams without nucleon exchange,
Fig~\ref{pd2.fig}.
These diagrams reproduce
the familiar Coulomb ladder diagrams representing interaction of two charged
particles with masses equal to $M_N$ and $M_d\approx2 M_N$.

The tree level diagram in
Fig.~\ref{pd2.fig} (a) is  proportional to $e^2/p^2$ which is consistent with
the power counting.
(b) is $\sim e^2/p^2\times (y^2 M_N^2)/(4\pi Q)$,
from the power counting. This is consistent with the actual calculation,
see Eqs.~\refeq{diagram_b} and~\refeq{diagram_b_exact} in
Appendix~\ref{powercounting}.
These two diagrams contribute the usual
tree level Coulomb pieces proportional to $1/p^2$. However,
Fig.~\ref{pd2.fig} (b) is bigger in the $Q/\Lambda$ counting.
We note that these diagrams with the deuteron wave function
renormalization factor $Z_d$ gives
the Coulomb potential in momentum space,
if we do a low-energy approximation of the
nucleon loop integral and keep only the LO contribution from the loop.
We have:
\beq
Z_d\times\[{\rm (a)+(b)}\]\Align i\frac{4\pi \alpha}{q^2}\equiv V_c(q)\ ,
\eeq
where $q$ is the photon momentum. It is reassuring to recover the familiar
result through the EFT power counting.

Fig.~\ref{pd2.fig} (c)$\sim \alpha^2 y^2 M_N^3/(p Q^3)$,
which looks odd since it scales as $1/p$. However, as we will show later it
does not contribute to the Coulomb modified amplitude, in perturbation.
With a little effort,
one concludes that a diagram similar to
Fig.~\ref{pd2.fig} (c) with $n>3$ photon propagators is infrared finite and
at most scale as $y^2 M_N/Q^2\times (\alpha M_N/Q)^n$, which is negligible
compared to the
infrared finite contributions from Fig.~\ref{PD1.fig}, and
such corrections are ignored. A $n=3$ photon (attached to a nucleon bubble)
diagram could
contribute as $y^2 M_N/Q^2\times (\alpha M_N/Q)^3\log(p/Q)\lesssim 0.02$ for
momentum $p\gtrsim 1$ MeV.
As we mention later, the numerical procedure used to
solve for the scattering amplitude does not yield reliable results below
momentum $p\approx 20$ MeV. Thus $\log(p/Q)$ corrections are ignored in the
present calculation as well.

From the power counting it follows that dressing any diagram by an extra
Coulomb photon attached to a nucleon bubble as in Fig.~\ref{pd2.fig} (b)
contributes a factor of $\alpha M_N/p$. Thus the diagram in
Fig.~\ref{pd2.fig} (d) $\sim e^2/p^2 (y^2 M_N^2)/(4\pi Q)\alpha M_N/p
=(b)\times \alpha M_N/p$, and (e) $\sim (a)\times \alpha M_N/p$. Dressing by an
extra photon attached to just a dibaryon field as in  Fig.~\ref{pd2.fig} (a)
contributes a factor of $\alpha M_N/p\times (4\pi Q)/(y^2 M_N^2)$. Thus these
are the effective range corrections to the Coulomb photons attached to the
nucleon bubble, as can be seen from  Fig.~\ref{pd2.fig} (f), etc.
See Appendix~\ref{powercounting} for more details. Finally,
dressing by two photons attached to the same nucleon bubble as in
Fig.~\ref{pd2.fig} (c) contributes a factor of
$\sim \alpha^2 M_N^2/Q^2\approx 0.02$. Thus Fig.~\ref{pd2.fig} (g) is a $2\%$
correction to (b), and (h) is a $2\%$ correction to (c), etc. Since we work to
only N$^2$LO in the purely strong interactions, which have an error of about
$3\%$,  we will ignore these $\approx 2\%$ electromagnetic effects.

\emph{To summarize}, the diagrams in
Fig.~\ref{pd2.fig} (a) and (b) are iterated to all order and they reproduce the
Coulomb ladder contribution scaling as $1/p^2$, $1/p^3$, etc.
We include Fig.~\ref{pd2.fig} (c), without iteration, which scales as $1/p$.
Iterating any diagram by two photons attached to a nucleon bubble as in
Fig.~\ref{pd2.fig} (c) only modifies the coefficient of the $1/p$, $1/p^2$,
$1/p^3$, etc., terms by about $2\%$ and we ignore such contributions. The
contributions from diagrams with $n>2$ photons attached to a single nucleon
bubble is infrared finite (except $n=3$) and negligibly small,
and we ignore such affects as well.

\begin{figure}[t]
 \begin{center}
  \epsfig{figure=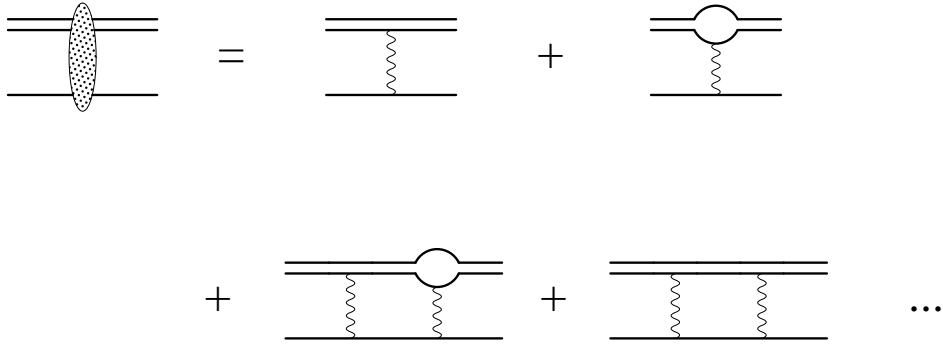, height=45mm}
 \end{center}
\caption{\it \protect The Coulomb scattering amplitude
$\mcal T_{c}$. The solid lines represent nucleons, double lines deuterons, wavy
lines photons. The deuteron-two-nucleon vertex coupling is $y$ and the
photon-nucleon, -deuteron coupling is $e$. }
\label{pc2.fig}
\end{figure}
We define the amputated Coulomb scattering amplitude by the diagrams in
 Fig.~\ref{pc2.fig} as:
\beq\eqn{C_Amp}
i\mcal T_{c}(\bfk,\bfp)\Align
i\frac{\alpha M_N^2 y^2}{2}\(\frac{1}{\gamma}-\rho\)
\frac{1}{({\bfk}-{\bfp})^2}\nonumber\\
&{}&+i2\pi\alpha M_N
\int {\frac{{\rm d}^3\bfq}{(2\pi)^3}\,}\frac{\mcal T_{c}(\bfk,\bfq)}
{-\gamma +\sqrt{3/4\bfq^2 -M_N E_T}-\frac{\rho}{2}\(3/4\bfq^2
-M_N E_T-\gamma^2\)}\nonumber\\
&{}&\hspace{1.2in} \times
\(\frac{1}{\gamma}-\rho\)
\frac{1}{\({\bfq}-{\bfp}\)^2}\ ,
\eeq
where the energy-momentum kinematics are the same as in Eq.~\refeq{S_Amp}.
In Eq.~\refeq{C_Amp}, we did not include the contribution from
 Fig.~\ref{pd2.fig} (c)
since for what we calculate later in Eq.~\refeq{phase_l}, contributions from
Fig.~\ref{pd2.fig} (c) in Eq.~\refeq{C_Amp} and Eq.~\refeq{Full_Amp} cancel in
 perturbation.

\end{subsection}
\begin{subsection}{ Coulomb with Strong Interaction}
\label{StrongAndCoulomb}

Most of what was said about the Coulomb photons also hold here. For example,
dressing the
strong scattering amplitude $T_s$ on either side by Coulomb photons,
as in Fig.~\ref{pd3.fig} (a), enhances
$T_s$ by factors of $\alpha M_N/p$.
However, there
are a couple of differences involving single photon exchange diagrams, as in Fig.~\ref{pd3.fig} (b) and (c).

From a naive power counting estimate Fig.~\ref{pd3.fig} (b)
$\sim (y^2 M_N/Q^2)\times \alpha M/Q$ which is infrared finite. It is as
large as the NLO strong interaction corrections to $T_s$
so one should include
it in the calculation.
However, a straightforward calculation shows that it is a $7\%$ effect.
We do not include such contributions in the calculation. Due to the absence
of this contribution in the present calculation, the theoretical error will be
around $7\%$.
The diagram in Fig.~\ref{pd3.fig} (c) is a bit more complicated.
Naively
this particular diagram gets equal sized contributions
from the $q\sim Q$ and the
$q\sim p$ part of the loop integration. Power counting indicates a
size $\sim 4\pi \alpha/Q^2$. However, a more careful analysis shows
(c) $\sim \alpha/Q^2 \log(p^2/Q^2)$ plus other negligibly small
constant pieces.
For $p\gtrsim 20$ MeV, neglecting the contribution from (c)$\lsim 1\%$ will
be consistent
with the other approximations.
We drop
contributions
from $q\sim Q$ but keep the contributions from $q\sim p$,
 for the diagrams in (c) and the similar
one with a photon attached to a nucleon bubble, for computational ease. This
approximation turns out to be valid when we compare our results with phase
shifts extracted from experimental data to within the accuracy assumed.
Incidentally, this approximation reduces to
iterating the strong interaction kernel with the
coordinate space Coulomb potential $V_c=\alpha/r$.

\begin{figure}[t]
 \begin{center}
  \epsfig{figure=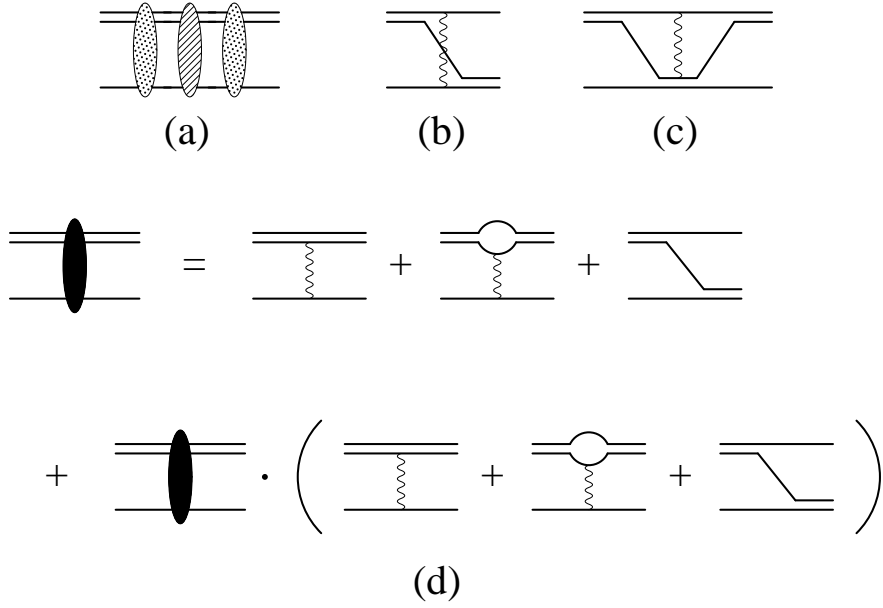, height=80mm}
 \end{center}
\caption{\it \protect The full scattering amplitude
$\mcal T_{Full}$. The solid lines represent nucleons, double lines deuterons,
 wavy
lines photons. The deuteron-two-nucleon vertex coupling is $y$ and the
photon-nucleon, -deuteron coupling is $e$.}
\label{pd3.fig}
\end{figure}
From Fig.~\ref{pd3.fig} (d), we get for the amputated scattering amplitude:
\beq\eqn{Full_Amp}
i\mcal T_{Full}(\bfk,\bfp)\Align
\frac{-i M_N y^2}{{\bfk}\cdot{\bfp}+\bfk^2+\bfp^2 -M_N E_T}
+i\frac{\alpha M_N^2 y^2}{2}\(\frac{1}{\gamma}-\rho\)
\frac{1}{({\bfk}-{\bfp})^2}\nonumber\\
&{}&-i\int {\frac{{\rm d}^3\bfq} {(2\pi)^3}\,}\frac{\mcal T_{Full}(\bfk,\bfq)}
{-\gamma +\sqrt{3/4\bfq^2 -M_N E_T}-\frac{\rho}{2}\(3/4\bfq^2
-M_N E_T-\gamma^2\)}\nonumber\\
&{}&\ \ \ \ \ \times\[ \frac{4\pi}{\({\bfq}+{\bfp}/2\)^2+3/4\bfp^2-M_N E_T}
-2\pi\alpha M_N\(\frac{1}{\gamma}-\rho\)\frac{1}{\({\bfq}-{\bfp}\)^2} \]
\nonumber\ ,\\
\eeq
 in the
presence of the strong and Coulomb effect,
with the approximations mentioned above.
The energy-momentum kinematics are the same as
in Eq.~\refeq{S_Amp}. Again we drop contribution from
Fig.~\ref{pd2.fig} (c) for the reason mentioned earlier,
following Eq.~\refeq{C_Amp}.

Note that diagrams similar to Fig.~\ref{pd3.fig} (c) with more than one photon
exchanges are infrared divergent. Such contributions are kept, though they
might be numerically small for momenta $p\gsim 40$ MeV. We include all
infrared divergent contributions from the Coulomb potential $V_c=\alpha/r$.

\end{subsection}
\end{section}
\begin{section}{Phase Shifts}
\label{phases}

Predicting the differential cross section for elastic $p$-$d$
scattering, from Eq.~\refeq{Full_Amp},
for direct comparison with experimental data involves
solving a multi-dimensional integral
equation. In this paper, to reduce the problem to
solving
a one-dimensional
integral equation, we calculate the Coulomb subtracted phase shifts
(after partial wave projections)
instead.
However, calculating phase shifts has the
disadvantage that it depends on precisely the definition used since it involves
subtracting some Coulomb effects from the full amplitude.
To compare our results with available phase shift analysis, we use the
same subtraction as conventionally defined including resummation of the
LO Coulomb effects which might not be necessary at momenta above say
$\sim 40$ MeV.
 The conventionally subtracted Coulomb effects in the phase shift
 analysis are the usual
Coulomb scattering amplitude with correction for the fact that the
charge of the deuteron
is not concentrated at its center but is bound to the position of the proton
in the deuteron~\cite{CSub}. 

The Coulomb modified
phase shift $\delta_l$ is defined as:
\beq\eqn{phase_l}
\delta^{(l)}(k)&\equiv&\delta^{(l)}_{Full}(k)-\delta^{(l)}_c(k)\ ,\nonumber\\
\delta^{(l)}_{Full}(k)\Align -i\frac{1}{2}
\log\[1+i\frac{2 M_N k}{3\pi}Z_d\mcal T^{(l)}_{Full}(k,k)\]\ ,\nonumber\\
\delta^{(l)}_{c}(k)\Align -i\frac{1}{2}
\log\[1+i\frac{2 M_N k}{3\pi}Z_d\mcal T^{(l)}_{c}(k,k)\]\ ,
\eeq
for every partial wave $l$. However, it is not possible to project out
 $\mcal T^{(l)}_{Full}(k,k)$ and $\mcal T^{(l)}_c(k,k)$
for any partial wave $l$ because
the Coulomb photon propagator is ill defined in the forward direction at any
momentum $k$.
This problem can be avoided following the proposal by Alt. et. al.,
see Ref.~\cite{alt} and the references there in for details.
The idea is simple: One
introduces a photon mass $\lambda$
as a regulator and calculates $\mcal T^{(l)}_{Full}$
and $\mcal T^{(l)}_c$ using any standard technique
for short range interaction. Then one numerically reduces the photon mass
$\lambda$  until the phase shift $\delta_l$ in
Eq.~\refeq{phase_l} which depends on the difference between
$\delta^{(l)}_{Full}$ and $\delta^{(l)}_c$, reaches a stable value.

The full scattering amplitudes with the photon mass $\lambda$ can now be
projected onto different partial waves and the $S$-wave amplitude is:
\beq\eqn{Full_Amp_l}
\mcal T^{(0)}_{Full}(k,p;\lambda)\Align
-\frac{2 M_N y^2}{k p}\frac{1}{2}Q\(\frac{k^2+p^2-M_N E_T}{k p}\)
    \nonumber\\
&{}&-\frac{\alpha M_N^2 y^2}{2 p k}\(\frac{1}{\gamma}-\rho\)\frac{1}{2}
Q\(-\frac{k^2+p^2+\lambda^2}{2 k p}\)\nonumber\\
&{}&-\int_0^\infty \mathrm{d} q\
\frac{\mcal T^{(0)}_{Full}(k,q;\lambda)}
{-\gamma +\sqrt{3/4q^2 -ME_T}-\frac{\rho}{2}\(3/4q^2
-M_NE_T-\gamma^2\)}\,\frac{q}{p}\nonumber\\
&{}&\ \ \ \ \ \times\[ \frac{2}{\pi}Q\(\frac{q^2+p^2-M_N E_T}{q p}\)
+\frac{\alpha M_N}{2\pi}\(\frac{1}{\gamma}-\rho\)
Q\(-\frac{q^2+p^2+\lambda^2}{2 q p}\)\]\nonumber\ ,\\
\eeq
with
\beq
Q(a)\Align\frac{1}{2}\int_{-1}^{1}\mathrm{d} x\frac{1}{x+a}\ .
\eeq

Similarly, the purely Coulomb
$S$-wave scattering amplitude with photon mass $\lambda$
is:
\beq\eqn{C_Amp_l}
\mcal T^{(0)}_{c}(k,p;\lambda)\Align
-\frac{\alpha M_N^2 y^2}{2 p k}\(\frac{1}{\gamma}-\rho\)\frac{1}{2}
Q\(-\frac{k^2+p^2+\lambda^2}{2 k p}\)\nonumber\\
&{}&-\int_0^\infty \mathrm{d} q\
\frac{\mcal T^{(0)}_{Full}(k,q;\lambda)}
{-\gamma +\sqrt{3/4q^2 -ME_T}-\frac{\rho}{2}\(3/4q^2
-M_NE_T-\gamma^2\)}\ \frac{q}{p}\nonumber\\
&{}&\hspace{1in}\times\frac{\alpha M_N}{2\pi}\(\frac{1}{\gamma}-\rho\)
Q\(-\frac{q^2+p^2+\lambda^2}{2 q p}\)\ .
\eeq

Using Eqs.~\refeq{phase_l},~\refeq{Full_Amp_l} and~\refeq{C_Amp_l}, we
 calculate the $S$-wave phase shifts. We find that at any given momentum $k$, a
 photon mass $\lambda$ in the range $k/10$ to $k/100$ gives a value that is
 numerically stable to about
within $1$-$3\%$ for $k\gtrsim 20$ MeV, with larger errors
for smaller momenta. This does not imply that the power counting is
 invalid below $20$ MeV. The sizes of the diagrams are still
 consistent with the power counting estimates. The numerical error is
 strictly
 associated with the partial wave decomposition. The screening photon
 mass method shows slow convergence to the $\lambda=0$ limit.
For momentum $k$ below $\sim 20$ MeV, our numerical
 routine does not converge in the numerical sense\footnote{Incidentally,
the authors in Ref.~\cite{BZ86} who also follow the screening photon mass
 procedure find significant numerical errors below momentum $k\sim 12$ MeV.
}.
 We also have a similar sized theoretical error:
higher order strong interaction effects such as
contribution of shape parameter $w_2$, etc., to the deuteron propagator in
Eq.~\refeq{deuteron} and Coulomb effects $\sim\alpha^2 M_N^2/Q^2\approx 0.02$
from diagrams such as those in
 Fig.~\ref{pd2.fig} (g).
 Therefore, we present phase
 shifts only for momenta $k\geq 20$ MeV. The results are shown in
Fig.~\ref{RealPhase.fig}, where we present the phase shift for $n$-$d$
 scattering as well.
For comparison, we also show results obtained
using potential model AV18~\cite{potPhase,tornowPhase}
and phase shift analysis obtained
from  experimental data~\cite{pdData}.
We note that numerical comparison of the
purely Coulomb phase shift
calculated in EFT and potential model~\cite{alt} agree to within $3\%$ for 
a given photon screening mass $\lambda$.

\makefigR{RealPhase.fig}{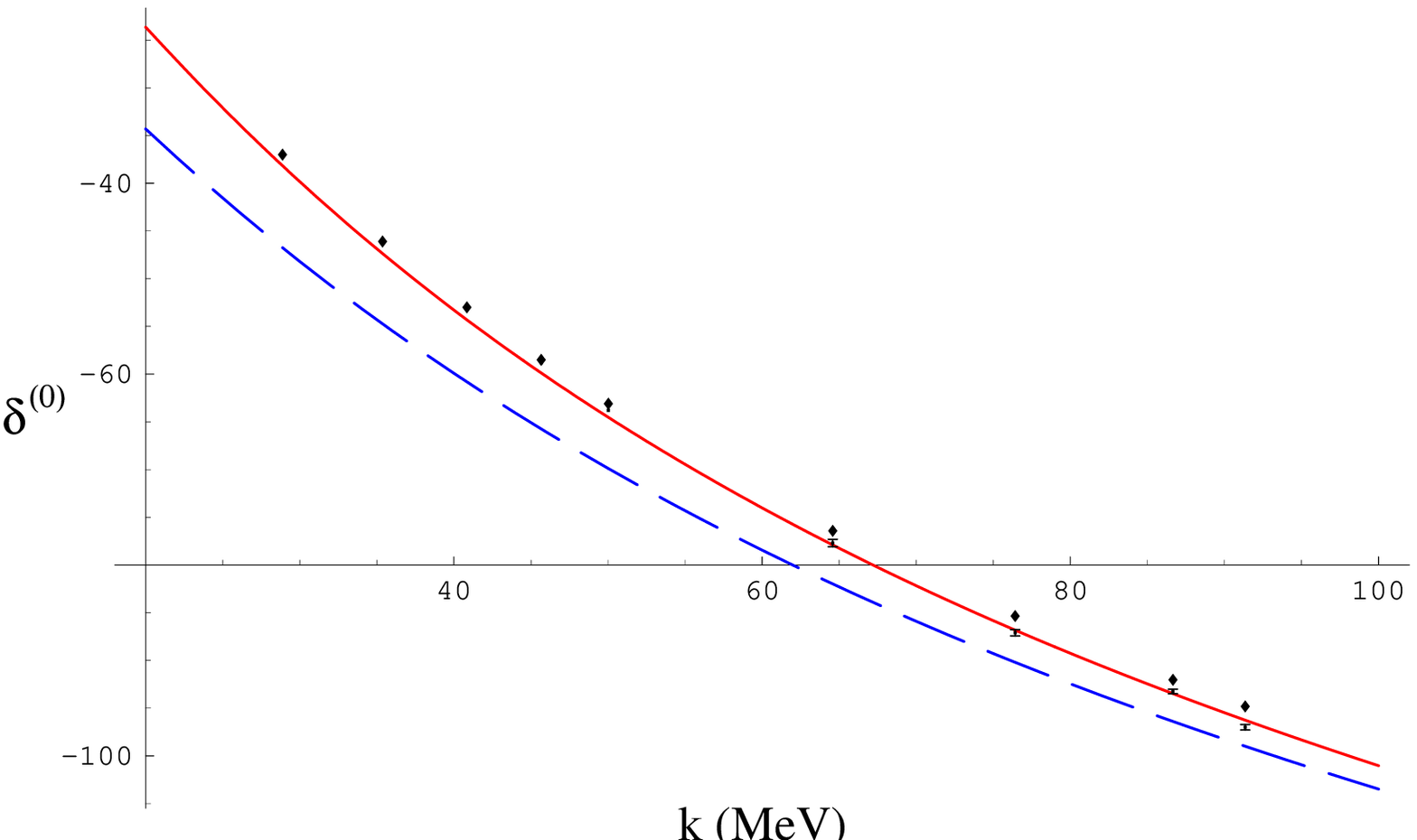}{3.2in}{\it \protect
Quartet $S$-wave $N$-$D$ phase shifts $\delta$ in degrees versus
center-of-mass momentum $k$ in MeV. Solid curve: EFT $p$-$d$ result;
dashed curve: EFT $n$-$d$ result; diamonds: potential model
results from Ref.~\cite{potPhase} below deuteron breakup and
from Ref.~\cite{tornowPhase} above breakup at momentum $k=52.77$ MeV;
dots with error bars:
TUNL phase shift analysis~\cite{pdData}. }
One can see some expected pattern from Fig.~\ref{RealPhase.fig}. Coulomb
effects are important at low-energies, accounting for as much as
$\approx 40\%$ difference between the $n$-$d$ and $p$-$d$ phase shifts at
momenta $k\sim 20$ MeV. The
  $40\%$ effect at $k\sim 20$ is in the Coulomb subtracted phase shift after
removing similar sized contributions from the total cross section.
It is reasonable to assume that at these low momenta the perturbation
has at least a significantly better rate of convergence after the resummation.
It would seem that resummation is unnecessary at higher
momenta $k\gsim 60$ MeV where the Coulomb effects in the subtracted
phase shift is as low as $6\%$. However, Coulomb effects would
still be large in the forward direction in the total cross section.
It is also evident that $p$-$d$
phase shift, calculated in EFT, potential model and TUNL phase shift analysis,
is consistently larger than the $n$-$d$ phase shifts.

Note that only certain quantities,
e.g. conventionally defined
$\delta^{l}=\delta^{(l)}_{Full}-\delta^{(l)}_{c}$, have a sensible
$\lambda\rightarrow 0$ limit~\cite{alt}. In order to ensure a sensible
$\lambda\rightarrow 0$ limit and to make a meaningful comparison with
available partial wave analysis, we use the conventional definition in
calculation $\delta^{(l)}$ including resumming Coulomb photons which
might not be necessary at large momenta.

The EFT result for $p$-$d$ phase shift agrees with the potential model result
and the TUNL phase shift analysis values, within the estimated $\sim7\%$
theoretical  and $\sim3\%$ numerical error.
On closer inspection, it appears that for
momenta $p\lesssim 60$ MeV, central value of EFT result is consistently
smaller than the TUNL values, whereas for very large momentum $k\sim 100$ MeV
it is slightly larger.
On the other hand the potential model results are
consistently larger by a similar amount.

\end{section}
\begin{section}{Summary}
\label{conclusion}
In conclusion, we developed a power counting for the Coulomb effects
in the quartet channel for $p$-$d$ systems in a low-energy pionless EFT. The
power counting reproduces $p$-$d$ scattering
results that have been known in potential models for
decades.
However, this power counting will be crucial for future EFT
three-nucleon calculations beyond the $p$-$d$ elastic scattering process, where
precision potential model results might not be available or well known.
The quartet channel $S$-wave elastic $p$-$d$ scattering phase shift was
calculated both below and above the deuteron breakup threshold.
 Calculations were performed up to
NLO in the Coulomb corrections and N$^2$LO in the strong interactions, with an
estimated theoretical
error of $\sim 7\%$ and $\sim 3\%$ respectively. An error of about $2\%$ was
found in the numerical evaluation of the scattering amplitude.
Within the estimated error, the EFT results agree with both potential
model calculations and phase shift analysis of experimental
data.

The LO and NLO Coulomb ladder
contribution is equivalent to contributions from the
coordinate space Coulomb potential $V_c=\alpha/r$.
The large $7\%$ error is primarily from the
single photon diagrams, similar to that in
Fig.~\ref{pd3.fig}(b) dressed with the LO scattering amplitude on the external
legs. These diagrams constitute the largest higher order corrections to the
Coulomb ladder.
There have been some discussion in the literature about Coulomb
polarization effects. These effects have been found to be negligible
for the present
calculation~\cite{BZ86} and they are not included.

The power counting for the Coulomb effects developed here could be applied
to higher partial waves in
both quartet and doublet channel for $p$-$d$ scattering.
As mentioned in the introduction, one can then go on to study
the so-called $p$-$d$ $A_y$ puzzle, and low-energy
many-body processes such as
$pd\rightarrow\gamma\ ^3$He, $dd\rightarrow n\ ^3$He, etc., that might be
important for placing
stricter bound on primordial light element abundances in
big-bang nucleosynthesis calculations~\cite{BNTT}. Triton beta decay is another
important low-energy process where three-body Coulomb effects are important.
This process is related to neutrino-deuteron scattering
and might have significance
to neutrino physics at SNO.

\end{section}
\begin{acknowledgments}
The authors thank P.~F.~Bedaque, D.~B.~Kaplan and M.~J.~Savage for various
useful discussions. We thank H.~W.~Hammer and D.~R.~Phillips for valuable
comments on the manuscript. W.~Schadow is thanked for bringing the work of
Alt. et. al. on screening photon masses to our attention.
We would also like to thank W. Tornow for making the TUNL phase shift
analysis data available.
We acknowledge support in part by the Natural Sciences and Engineering Research
Council of Canada.

\end{acknowledgments}
\appendix
\begin{section}{The deuteron propagator with Coulomb photon}
\label{powercounting}
To understand the power counting rules of Section~\ref{Coulomb}, it
is most convenient to start with the leading order deuteron
propagator, without the range correction:
\beq
i\Delta(p_0,\bfp)\Align -i \frac{4\pi}{M_N y^2}\frac{1}
{-\gamma +\sqrt{\bfp^2/4-M_N p_0-i\epsilon}
 }\nonumber\\
\Align -i \frac{4\pi}{M_N y^2}
\frac{\gamma+\sqrt{\bfp^2/4-M_N p_0}}{\bfp^2/4-M_N p_0-\gamma^2-i\epsilon}
\ .
\eeq

In elastic $p$-$d$ scattering, every Feynman diagram with a deuteron
propagator involves only a single nucleon propagator at equal times,
 see Figs.~\ref{PD1.fig},\ref{pd2.fig},\ref{pc2.fig} and~\ref{pd3.fig}.
The deuteron propagator momentum in any diagram can be written such that
$(p_0=E_d+q_0,\bfp=\bfq)$ with the corresponding nucleon line carrying
energy-momentum
$(E_N-q_0,-\bfq)$ where $E_d=p^2/(4 M_N)-\gamma^2/M_N$ is the
incoming/outgoing deuteron energy and $E_N=p^2/(2M_N)$ is the nucleon
incoming/outgoing energy with incoming/outgoing momenta $p$,
for a generic loop momentum
$(q_0,\bfq)$. We get:
\beq
i\Delta(E_d+q_0,\bfq)\Align
 -i \frac{4\pi}{M_N y^2}
\frac{\gamma+\sqrt{\frac{\bfq^2-\bfp^2}{4}-M_N q_0+\gamma^2}}
{\frac{\bfq^2-\bfp^2}{4}-M_N q_0-i\epsilon}\ ,
\eeq
where the factors of $\gamma^2$ cancel in the
deuteron pole.
Carrying out the $q_0$ integral, as in Eq.~\refeq{diagram_f},
 and putting the nucleon on-shell with
$q_0=E_N-\bfq^2/(2 M_N)$ gives
\beq
i\Delta(E_d+E_N-\frac{\bfq^2}{2 M_N},\bfq)\Align
 -i \frac{4\pi}{M_N y^2}
\frac{\gamma+\sqrt{3\frac{\bfq^2-\bfp^2}{4}+\gamma^2}}
{3\frac{\bfq^2-\bfp^2}{4}}\ .
\eeq
Including the effective range $\rho$ is straightforward, and we get:
\beq
i\Delta(E_d+E_N-\frac{\bfq^2}{2 M_N},\bfq)\Align
 -i \frac{4\pi}{M_N y^2}
\frac{\gamma+\sqrt{3\frac{\bfq^2-\bfp^2}{4}+\gamma^2}}
{3\frac{\bfq^2-\bfp^2}{4}}
\frac{1}{1-\frac{\rho}{2}
\(\gamma+\sqrt{3\frac{\bfq^2-\bfp^2}{4}+\gamma^2}\)}\ .
\eeq
We will ignore the effective range corrections $\sim\rho\gamma\sim Q/\Lambda$
 in the
following discussion for simplicity.
For small momenta $p\ll\gamma\sim Q$, when $q\sim p$ ($d^3 q\sim p^3$) we
get $\Delta\sim (4\pi)/(M_Ny^2) \gamma/p^2\sim Q/p^2$ whereas when
 $q\sim \gamma$ ($d^3 q\sim \gamma^3$)
 we get $\Delta \sim (4\pi)/(M_Ny^2 \gamma)\sim 1/Q$.
This scaling of the deuteron
propagator is far from obvious. Even though one might naively expect
$\Delta\sim 1/Q$ from dimensional analysis in
Eqs.~\refeq{deuteronLO} and~\refeq{deuteron},
this is invalidated since
the deuteron can be put on shell which exactly cancels the factors of
$\gamma^2$ from the deuteron pole.

Now, even though the contribution from the deuteron pole is enhanced
for $q\sim p$, its contribution is typically suppressed by
$d^3 q\sim p^3$ except when the loop integral also involves a
Coulomb photon propagator which scales as $1/q^2\sim 1/p^2$. This
explains why there are no $1/p$ infrared enhancements in the diagrams
without photons in
Fig.~\ref{PD1.fig}, where the dominant contribution comes from loop
momentum $q\sim \gamma\sim Q$ with $\Delta\sim 1/Q$. On the other hand
diagrams with Coulomb photons get infrared enhancements from $q\sim p$, 
and $\Delta\sim Q/p^2$.
For example, from Fig.~\ref{pd2.fig} (f) we get the contribution
(ignoring factors of $i$,$2$, $\pi$, etc.,):
\beq\eqn{diagram_f}
&\sim&\int\frac{{\rm d}q_0}{2\pi} {\frac{{\rm d}^3\bfq}{(2\pi)^3}\,}
\Delta(E_d+q_0,\bfq)
\frac{1}{-q_0+E_N-\frac{\bfq^2}{2M_N}}
\frac{\alpha}{\(\bfq-\bfp\)^2} \frac{\alpha}{\(\bfq-\bfk\)^2}\nonumber\\
&\sim&\int {\frac{{\rm d}^3\bfq}{(2\pi)^3}\,}
\Delta(E_d+E_n-\frac{\bfq^2}{2M_N},\bfq)
\frac{\alpha}{\(\bfq-\bfp\)^2}
\frac{\alpha}{\(\bfq-\bfk\)^2}\nonumber\\
&\sim&\frac{\alpha^2}{M y^2}\int {\frac{{\rm d}^3\bfq}{(2\pi)^3}\,}
\frac{(\gamma+\sqrt{3\bfq^2/4-3\bfp^2/4+\gamma^2})}
{3\bfq^2/4-3\bfp^2/4}
\frac{1}{\(\bfq-\bfp\)^2}
\frac{1}{\(\bfq^2-\bfk\)}\ ,
\eeq
where the deuteron energy $E_d=\bfp^2/(4M_N)-\gamma^2/M_N$ and
the nucleon energy
$E_N=\bfp^2/(2M_N)$ with momenta $\bfp$, $\bfk$ defined as
in Eq.~\refeq{S_Amp}. This integral is dominated by momenta $q\sim |p|=|k|$ and
the contribution is, without wave function renormalization,
\beq
&\sim& \frac{\alpha}{p^2}\frac{\alpha Q}{M y^2 p}
= \frac{\alpha}{p^2}\frac{\alpha M_N}{p}\times\frac{Q}{M_N^2 y^2}\ .
\eeq

Diagrams without a deuteron propagator involve three nucleons. After
carrying out the integral over the energy component $q_0$ and putting
a nucleon on-shell, one is left with two nucleons that are always
regulated in the infra red by the
deuteron binding momentum $\gamma$, as these involve two nucleons that
carry the energy $E_d$ of the ``parent'' deuteron. For example, in
Fig.~\ref{pd2.fig} (b) we have
\beq\eqn{diagram_b}
&\sim& y^2\int \frac{{\rm d}q_0}{2\pi} {\frac{{\rm d}^3\bfq}{(2\pi)^3}\,}
\frac{1}{-q_0-\frac{\bfq^2}{2M_N}}\frac{1}{q_0+E_d-\frac{(\bfq+\bfp)^2}{2M_N}}
\frac{1}{q_0+E_d-\frac{(\bfq+\bfk)^2}{2M_N}}
\frac{\alpha}{(\bfp-\bfk)^2}\nonumber\\
&\sim& \frac{\alpha y^2 M_N^2}{(\bfp-\bfk)^2}\int {\frac{{\rm
      d}^3\bfq}{(2\pi)^3}\,}
\frac{1}{\frac{(\bfq+\bfp)^2}{2}+\frac{\bfq^2}{2}-\frac{\bfp^2}{2}+\gamma^2}
\frac{1}{\frac{(\bfq+\bfk)^2}{2}+\frac{\bfq^2}{2}-\frac{\bfp^2}{2}+\gamma^2}\nonumber\\
&\sim& \frac{\alpha y^2 M_N^2}{\bfp^2 Q}\ ,
\eeq
where the nucleon propagators get regulated in the infrared
by the deuteron binding
momentum $\gamma\sim Q$ with $|\bfp|=|\bfk|$.
After wave function
renormalization $Z_d\sim
Q/(M_N^2 y^2)$, it gives a contribution $\sim \alpha/p^2$.
Exact evaluation of the integral
gives
\beq\eqn{diagram_b_exact}
\frac{1}{2\pi(\bfp-\bfk)^2}
\arctan\[\frac{1}{2}\sqrt{\frac{(\bfp-\bfk)^2}{4\gamma^2-\bfp^2}}\]\sim
\frac{1}{8\pi\gamma}+\mcal O(p^2),
\eeq
which is
consistent with the power counting estimate (including all the
numerical factors).

 \emph{To summarize}, all the nucleon propagators scale as $M_N/Q^2$, the
 deuteron propagator scales as $Q/q^2$ and the Coulomb photon
 propagators scale as $1/q^2$. The loop momentum scales as either
 $q\sim p\ll Q$ or as $q\sim Q$, and correspondingly the deuteron
 propagator scales as $1/Q$ or $Q/p^2$. In estimating the dominant
contributions from
 a loop integral for $q\sim p$, one has to consider the suppression
 factors from $d^3 q\sim p^3$ along with infrared enhancements
 from the deuteron $Q/p^2$ and the photons $1/p^2$.

\end{section}

\end{document}